\documentclass[fleqn,10pt]{wlscirep}
\title{Network-based indicators of Bitcoin bubbles}

\newcommand\iac[1]{#1}

\author[1,2]{Alexandre Bovet}
\author[3]{Carlo Campajola}
\author[4]{Jorge F. Lazo}
\author[5]{Francesco Mottes}
\author[6]{Iacopo Pozzana}
\author[7]{Valerio Restocchi}
\author[8]{Pietro Saggese}
\author[8]{Nicol\'o Vallarano}
\author[8*]{Tiziano Squartini}
\author[9]{Claudio J.\ Tessone}
\affil[1]{naXys, University of Namur, Rempart de la Vierge 8, B-5000 Namur, Belgium}
\affil[2]{ICTEAM, Université catholique de Louvain, Avenue George Lemaître 4, B-1348 Louvain-la-Neuve, Belgium}
\affil[3]{Scuola Normale Superiore, I-56126 Pisa, Italy}
\affil[4]{Lund University, Professorsgatan 1,  Lund, Sweden}
\affil[5]{Universit\`a di Torino, Via Pietro Giuria 1, I-10125 Torino, Italy}
\affil[6]{Birkbeck - University of London, Malet Street, WC1E 7HX London, UK}
\affil[7]{ECS, University of Southampton, SO17 1BJ Southampton, UK}
\affil[8]{IMT School for Advanced Studies Lucca, I-55100  Lucca, Italy}
\affil[9]{URPP Social Networks, University of Zurich, CH-8050 Z\"urich, Switzerland}

\affil[*]{tiziano.squartini@imtlucca.it}


\begin{abstract}

The functioning of the cryptocurrency Bitcoin relies on the open availability of  
the entire history of its transactions.
This makes it a particularly interesting socio-economic system to analyse from the point of view of network science.
Here we analyse the evolution of the network of Bitcoin transactions between users. We achieve this by using the complete transaction history from December 5th 2011 to December 23rd 2013. This period includes three bubbles experienced by the Bitcoin price. In particular, we focus on the global and local structural properties of the user network and their variation in relation to the different period of price surge and decline. By analysing the temporal variation of the heterogeneity of the connectivity patterns we gain insights on the different mechanisms that take place during bubbles, and find that hubs (i.e., the most connected nodes) had a fundamental role in triggering the burst of the second bubble. Finally, we examine the local topological structures of interactions between users, we discover that the relative frequency of triadic interactions experiences a strong change before, during and after a bubble, and suggest that the importance of the hubs grows during the bubble. These results provide further evidence that the behaviour of the hubs during bubbles significantly increases the systemic risk of the Bitcoin network, and discuss the implications on public policy interventions.

\end{abstract}

\begin{document}

\flushbottom
\maketitle

\thispagestyle{empty}


\section*{Introduction}
Designed under the pseudonymous name of Satoshi Nakamoto, and introduced by a disruptive paper in 2008\cite{nakamoto2008bitcoin} while the world was challenged by the aftermaths of the financial crisis, Bitcoin is in essence a series of cryptographical protocols that solve the double-spending problem, i.e. prevent the same digital token from being spent more than once, in the absence of a third party that verifies and guarantees the validity of transactions.
More in detail, Bitcoin consists of a decentralized peer-to-peer network, composed by users that transact bitcoins among them; once it is validated by a network of miners according to the consensus rules that are part of the protocol, these transactions are included in a public and distributed transactional database, the blockchain ledger\cite{halaburda2016beyond, antonopoulos2017mastering,glaser2017pervasive}.
Few years after the date of its release, this digital currency has showed to be able to attract an increasing number of users, both because of speculative reasons\cite{gandal2018price,chu2015statistical,hayes2017cryptocurrency}, and  because of the trust of early adopters in the potentialities of this innovative technology\cite{bohme2015bitcoin,gervais2014bitcoin}.
In fact, the number of users and ergo the number of transactions within the bitcoin network has witnessed a remarkable burst which also has lead to an increment on its value in the market and consequently to some price bubbles and respective crashes\cite{Garcia:2014,wheatley2018bitcoin,gerlach2018dissection}; on the other hand, the novelties introduced by the bitcoin protocol have allowed a numerous number of innovative analyses and make the bitcoin network a particularly interesting case of study\cite{foley2018sex}.
Indeed, a remarkable feature of the transaction verification mechanism on which bitcoin system relies on is that the transaction history since the creation of the currency is openly accessible.
The availability of the complete transaction history allows to investigate the structural properties of the network of bitcoin users and to examine their relations
with its different growing phases.
The structure and dynamics of the network of bitcoin users has only recently started to be investigated.
Looking at the network of transactions between addresses, Kondor et al.\cite{Kondor:2014} have shown that the in-degree distribution of nodes, i.e. the number of incoming transactions of nodes, relates with nodes wealth distribution.
Parino et al.\cite{parino2018analysis} have investigated the network of the international bitcoin flow to identify socio-economic factors driving its adoption by country.\\

Here, we reconstruct the network of transaction between users by merging addresses apparently owned by different users making it closer to reality than the raw network of addresses\cite{Tessone:2018,meiklejohn2013fistful}.
We study the evolution of the network global quantities, such as the variation coefficient of the degree distributions, the sizes of the largest strongly and weakly connected components and 
we also investigate the evolution of the local structure of the networks by 
examining so-called \textit{network motifs}\cite{milo2002network}.
Network motif, defined as statistically recurrent subgraphs, were shown to implement simple functionalities that contribute to the complex behaviour of the system as a whole \cite{milo2002network}. This modular organization at the local scale was also shown to be common to a wide range of real networks \cite{milo2004superfamilies}.
In particular, the abundance of certain triadic motifs has been identified as a early-warning signal for topological collapse of inter-banking networks\cite{Squartini:2013}.

\section*{Results}

\subsection*{Evolution of global network measures}

To be able to detect patterns occurring at different time scales, we construct two time series of networks, using a integration time of one day and one week respectively. We refer to Wheatley et al. 2018.\cite{wheatley2018bitcoin}\ to identify the start and end date of the various bubbles and focus on the three first bubbles, reported in Table \ref{bubdates}. Our observation period starts 5 months before the onset of the first bubble, namely on December 5th 2011, and finishes on December 23rd 2013, one month after the burst of the last bubble taken into account. 

\begin{table}
\centering
\begin{tabular}{|c|c|c|}
\hline
Bubble&	Start		&	End	\\
\hline
1	&	2012-05-25	&	2012-08-18	\\
2	&	2013-01-03	&	2013-04-11	\\
3	&	2013-10-07	&	2013-11-23	\\
\hline
\end{tabular}
\caption{Time intervals of the three Bitcoin bubbles occurring between May 2012 and 
January 2014\cite{wheatley2018bitcoin}.}
\label{bubdates}
\end{table}


We explored the evolution of several network properties that can be relevant to highlight events that might be related to irrational exuberance of agents in the market or to crisis triggering phenomena such as liquidity imbalances. The following measures are defined:
\begin{itemize}

\item Number of nodes (wallets) in the network
Bubbles are characterised by a rising number of active users, especially at the end of the critical period.
\item Size of largest strongly and weakly Connected Components.
\item Ratio between the (total, in/out) degree of the most connected and second most connected node, dubbed as ``Degree Gap Ratio".
\item Ratio between the in and out degree of the most connected node, dubbed as ``Hub in/out Degree Ratio".
\end{itemize}

\begin{figure}
\centering
\includegraphics[width=.95\linewidth]{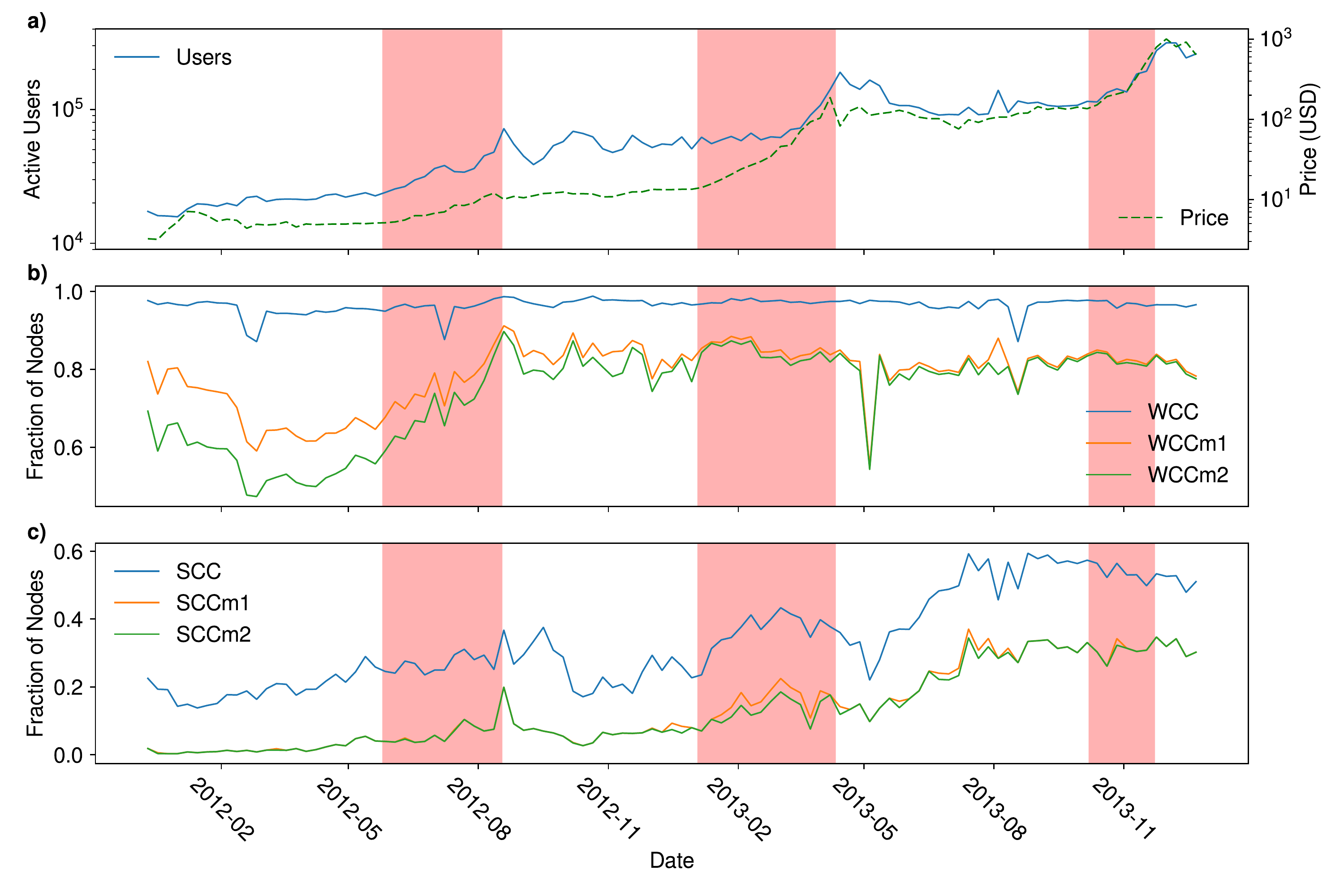}
\caption{\textbf{Bitcoin network and price evolution}.
The measures are performed on networks reconstructed each week.
The three bubbles are indicated by a shaded area (from onset until burst day).
a) Number of active users, i.e. number of nodes in the networks, as a function of time
(blue) and Bitcoin price in USD (green).
b) Size of the largest weakly connected component (WCC) of the full network
divided by the total number of nodes (blue),
largest weakly connected component after the removal of the highest degree node (WCCm1, orange) and the removal of the two nodes with highest degree (WCCm2, green).
c) Similar measure than in b) but using the largest strongly connected component 
instead of the largest weakly connected component.}
\label{fig:cc_users}
\end{figure}

As a first step we analyse the time series of the number of active users alongside the price evolution (Fig \ref{fig:cc_users}a), aggregated at the weekly time scale, which shows the strong correlation between the two, especially during bubble periods. This is easily explained by the herding feedback mechanism \cite{Garcia:2014}, where an initial price hike is followed by an increased popularity of the asset that lures more users into the market. The increase in activity of typically buying users (since they are getting into the market) further boosts the growth of price which eventually grows at unsustainable rate and collapses, either because of growth slowdown that triggers speculators to sell or after some negative news release that leads to panic sales.\\
Next we try to quantify how central are the most connected nodes in the circulation of money and whether their centrality in the network of transactions is related or not to the price surge and fall. To do so, we compute the size of the largest Strongly Connected Component (SCC) and Weakly Connected Component (WCC), namely the largest set of vertices that have a path running between them in both directions, for the SCC, and in at least one direction for the WCC. We make this analysis on the full network and on the networks where the most connected (SCCm1, WCCm1) and the two most connected nodes (SCCm2, WCCm2) are removed from the network with all of their edges. This would highlight the role of the largest and second largest hubs as intermediaries in the market. In particular the WCC is relevant to show if there are some fractions of the economy that never interact with others or that do so only through one of the two hubs, while the SCC can be an indicator of how efficiently a bitcoin can pass from one end to the other of the network and how this happens thanks to hubs. We plot these indicators in Fig \ref{fig:cc_users}b, \ref{fig:cc_users}c. We see how the vast majority of users are always connected to the network since the largest WCC includes almost all nodes. However, the importance of the first hub changes throughout the time period we consider, decreasing in time and in particular during the first bubble. We see that the first bubble eliminates a second important hub (WCCm1), and this could be related to the unmasking of the Ponzi scheme behind Bitcoins Savings \& Trust \cite{wheatley2018bitcoin} who might have been that player. After that event, the centrality of the first hub stabilises in the WCC and no other node has similar importance in that almost 20\% of nodes are connected to the market only through it. \\
Similar conclusions on the centrality of the first hub come from the gap between the SCC and SCCm1 which is again close to 20\% of the total nodes in the network. The circulation of money in the network rises as time goes on, and after the first bubble the first hub, although very central, is not any more the only passage point. This is a signal that smaller nodes start trading among themselves without relying only on the intermediation of the large exchange (likely Mt. Gox) and the velocity of money through the network is not completely hindered if the first hub goes in distress thanks to these secondary channels. It is interesting to notice that this doesn't happen at the expense of the importance of the first hub, which is indeed increasing its centrality, and neither through the formation of a secondary hub, since the second largest node is irrelevant to the size of the SCC and WCC. \\
However, we do not find consistent signals fore-running bubble onsets or bursts in these global measures. This could be related to the developing nature of the market, which is in constant evolution and for which these global features undergo such big changes that even comparing between time frames is extremely hard. We thus turn to more microscopic measures, and expose the results in the next section.

\begin{figure}
\centering
\includegraphics[width=.95\linewidth]{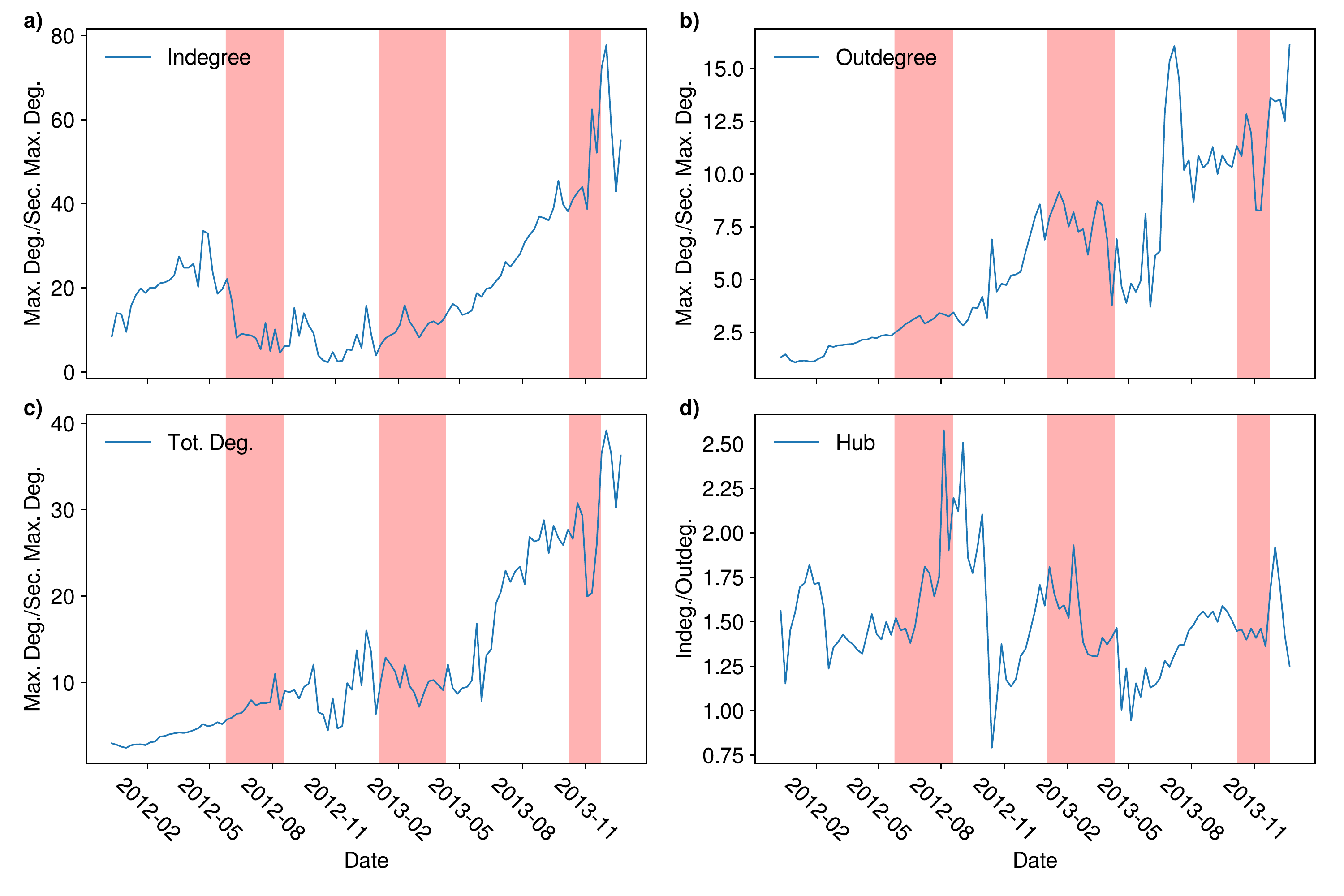}
\caption{\textbf{Degree gap ratios of the two largest hubs}.
The three bubbles are indicated by shaded area (from onset until burst day).
a) Ratio of the in-degree of the two nodes with highest in-degree.
b) Ratio of the out-degree of the two nodes with highest out-degree.
c) Ratio of the total degree of the two nodes with highest total degree.
d) Ratio of in-degree over out-degree for the largest hub.}
\label{fig:degrees}
\end{figure}

\subsection*{Heterogeneity measures}

\begin{figure}
\centering
\includegraphics[width=.75\linewidth]{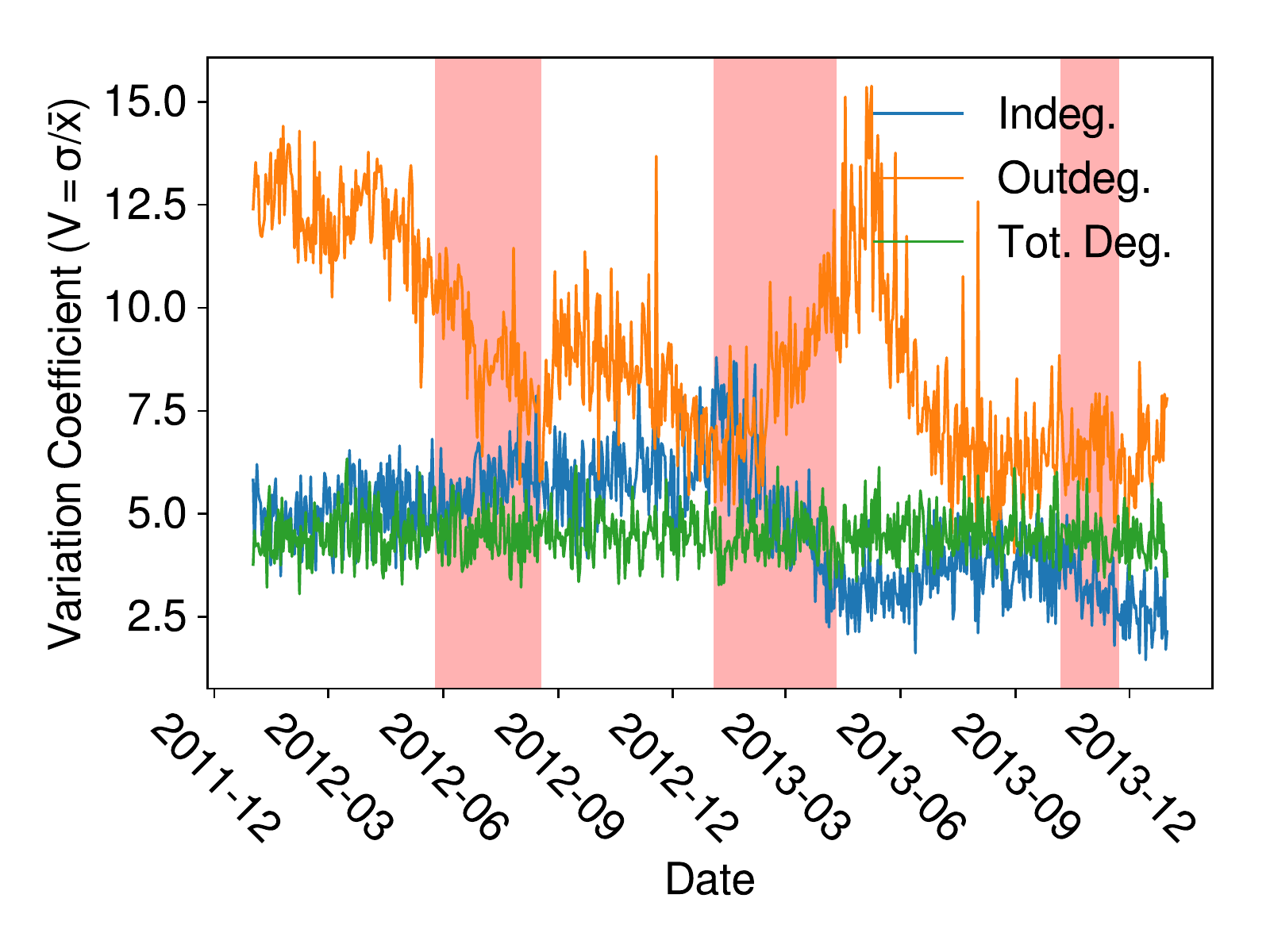}
\caption{\textbf{Heterogeneity evolution of the degree distributions}.
Variation coefficient of the in-degree (blue), out-degree (orange)
and total degree (green) distributions of daily networks.
The bubbles are indicated by the pink shaded areas (from onset until burst day).
}
\label{fig:var_coeffs}
\end{figure}


In this section we examine several measures of heterogeneity for three distributions, namely the in-degree, out-degree and total degree.
Specifically, we use a daily aggregation windows for network and cover two years of transactions, from January 2012 to December 31st 2013.
For each type of network and distribution, we compute the variation coefficient, i.e., $V=\frac{\sigma}{\bar{x}}$, where $\sigma$ and $\bar{x}$ are the standard deviation and the average of the empirical distribution, respectively.
As all the distributions display heavy tails, this measure of heterogeneity may depends on the size of the networks.
To account for the varying size of the networks at every point in time, we build the distributions only considering $N_{sample}=1784$ nodes for each daily network, where $N_{sample}$ is the number of nodes the smallest network possesses, and measure the average value of $V$ over 100 random sample for each distribution.
Our results, displayed in Figure \ref{fig:var_coeffs}, show that during the second bubble, which occurred between January 3rd 2013 and April 11th 2013, the in-degree distribution becomes far more homogeneous than before the bubble, whereas the out-degree distribution exhibit a surge of heterogeneity.

Fig. \ref{fig:degrees} shows the relative degrees of the largest and second largest hub, as well as the out-degree-in-degree ratio of the largest hub. Specifically, the latter measure suggests that the largest hub lowers the number of people it buys from with respect to the number of people it sells to, which is expected during a bubble, since more and more low-degree nodes enter the market following the price surge (see Fig. \ref{fig:cc_users}). Interestingly, figures \ref{fig:degrees} (a) and \ref{fig:degrees} (b) show that the the in-degree (out-degree) of the largest hub grows (decreases) with respect to that of the second largest hub, suggesting that the second largest hub follows a similar dynamics to that of the largest hub, but to a greater extent. Indeed, these results suggest that, during the second bubble, the second largest hub increases the number of customers it sells to, whereas it lowers the number of customers it buys from. It is worth noting that this is purely a structural change, since the largest hub keeps a null trade balance throughout this period. 

These structural changes are consistent with the changes in the heterogeneity of the in-degree and out-degree distributions, and suggest that there are two hubs that centralise the market by selling bitcoins to most of the traders that enter the market during the bubble, resulting in a significant increase of the systemic risk. Indeed, if only a few hubs account for most of the transactions in the network, if at any point in time one of them fails, the whole network may crash. This is exactly what happened on April 10th 2013, when Mt Gox, the major Bitcoin exchange, broke under the high trading volume, triggering the burst of the bubble.


%

\subsection*{Triadic motifs analysis}

Triadic motifs, i.e. all the possible directed patterns connecting three vertices, are the natural generalizations of directed clustering coefficients and the starting point for the understanding of a complex network self-organization in communities. Thirteen, non-isomorphic, triadic directed patterns (reported in Fig. \ref{fig:motifs}) can be identified and classified. Given a real, binary, directed matrix $\mathbf{A}$, the motifs occurrences $N_m$ can be written as reported in table \ref{motable}. 

\begin{table}
\centering
\begin{tabular}{c|c}
\hline
Motif $m$ &	$N_m$ \\
\hline
1 & $\sum_{i\neq j\neq k}(1-a_{ij})a_{ji}a_{jk}(1-a_{kj})(1-a_{ik})(1-a_{ki})$\\
2 & $\sum_{i\neq j\neq k} a_{ij}(1-a_{ji})a_{jk}(1-a_{kj})(1-a_{ik})(1-a_{ki})$\\
3 & $\sum_{i\neq j\neq k} a_{ij}a_{ji}a_{jk}(1-a_{kj})(1-a_{ik})(1-a_{ki})$\\
4 & $\sum_{i\neq j\neq k} (1-a_{ij})(1-a_{ji})a_{jk}(1-a_{kj})a_{ik}(1-a_{ki})$\\
5 & $\sum_{i\neq j\neq k} (1-a_{ij})a_{ji}a_{jk}(1-a_{kj})a_{ik}(1-a_{ki})$\\
6 & $\sum_{i\neq j\neq k} a_{ij}a_{ji}a_{jk}(1-a_{kj})a_{ik}(1-a_{ki})$\\
7 & $\sum_{i\neq j\neq k} a_{ij}a_{ji}(1-a_{jk})a_{kj}(1-a_{ik})(1-a_{ki})$\\
8 & $\sum_{i\neq j\neq k} a_{ij}a_{ji}a_{jk}a_{kj}(1-a_{ik})(1-a_{ki})$\\
9 & $\sum_{i\neq j\neq k} (1-a_{ij})a_{ji}(1-a_{jk})a_{kj}a_{ik}(1-a_{ki})$\\
10 & $\sum_{i\neq j\neq k} (1-a_{ij})a_{ji}a_{jk}a_{kj}a_{ik}(1-a_{ki})$\\
11 & $\sum_{i\neq j\neq k} a_{ij}(1-a_{ji})a_{jk}a_{kj}a_{ik}(1-a_{ki})$\\
12 & $\sum_{i\neq j\neq k} a_{ij}a_{ji}a_{jk}a_{kj}a_{ik}(1-a_{ki})$\\
13 & $\sum_{i\neq j\neq k} a_{ij}a_{ji}a_{jk}a_{kj}a_{ik}a_{ki}$\\
\hline
\end{tabular}
\caption{Classification and definitions of the triadic motifs abundances.}
\label{motable}
\end{table}

\begin{figure}
\centering
\includegraphics[width=.95\linewidth]{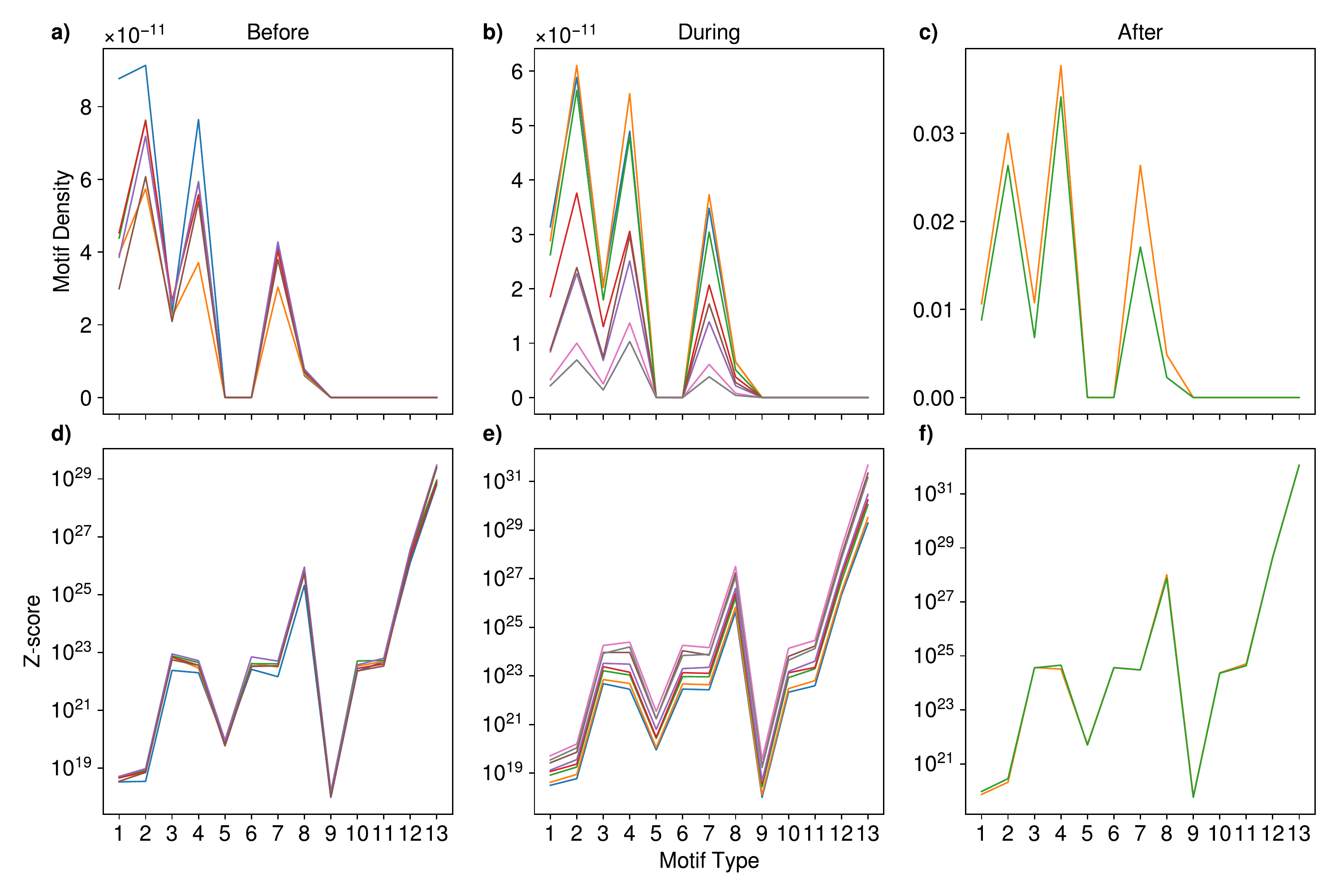}
\caption{\textbf{Frequency evolution of the 13 triadic motifs}.
}
\label{fig:motifs_frequency}
\end{figure}

The three upper panels of Figure \ref{fig:motifs_frequency} show the density of the 13 triadic motifs during three periods that have been chosen to monitor the system \emph{before}, \emph{during} and \emph{after} the bubble happening in the period of time from December 5th 2011 until September 18th 2012, taken on weekly aggregated networks. The three lower panels of Figure \ref{fig:motifs_frequency} show the z-scores of the same motifs.

Upon visually inspecting the three upper panels, it is apparent that the bubble is indeed characterised by motifs profiles that differ from both the previous and the following period. It is apparent that the same motifs (i.e. 1, 3, 6) are overrepresented during the three periods: upon inspecting these three kinds of motifs, we see that they are constituted by a basic unit of two non-reciprocated dyads. We suspect these to characterise the topological structure of the hubs, reflecting a huge selling activity in all periods. This is particularly evident, however, when considering the pre-bubble period from December 5th 2011 until May 25th 2012: this evidence leads us to assume that monitoring the hubs connectivity may be useful to detect upcoming critical activity.

These structural changes are consistent with the changes in the heterogeneity of the in-degree and out-degree distributions, and suggest that there are two hubs that centralise the market by selling bitcoins to most of the traders that enter the market during the bubble, resulting in a significant increase of the systemic risk. Indeed, if only a few hubs account for most of the transactions in the network, if at any point in time one of them fails, the whole network may crash. 

\section*{Conclusions}



In this paper we analyse the impact of  structural properties of the Bitcoin transaction network on the generation and  crash of bubbles in the exchange with respect to fiat currencies. Specifically, we examine network features such as heterogeneity of the degree distributions and frequency of connectivity patterns (i.e., motifs). We find significant changes in these properties during the period of price bubbles. A more detailed analysis unveils that, during the first bubble, the frequency of motifs indicating the relationship hubs have with new, low-degree users changes significantly; this suggests that hubs take an even more important role in becoming liquidity providers. These results are confirmed in the second bubble: There, by analysing the heterogeneity of the in-degree, out-degree, and total degree distributions, we find that there is a significant widening (narrowing) of the out-degree (in-degree) distributions, whereas the total degree does not change its distribution significantly. 
 By performing additional analyses on the two largest hubs, we find that these structural changes - similar to what is observed during the first bubble - is likely to be caused by the \emph{centralising} role hubs take on as liquidity providers. 

Although we find that measures can explain well some price bubbles but not others, these results highlight that tracking properties of hubs in the transaction network is key for understanding the underlying mechanisms of a bubble. Moreover, at least in the first three Bitcoin bubbles, the behaviour of hubs significantly increased the systemic risk of the Bitcoin economy, eventually leading to systemic failure and sudden price crashes. 

These results also suggest that Bitcoin bubbles are difficult to forecast, but can be prevented, or at least alleviated, by introducing policies that aim at reducing the importance of large hubs in the network. In future work, we plan to extend our analysis by introducing new structural measures and by covering all the bubbles that happened to date. 

\section*{Methods}

\subsection*{User network reconstruction}

An element of the Bitcoin protocol is that it attempts to preserve anonymity of users in a way that is better defined as \textit{pseudonimity}: transactions take place without the need of a third party and users cannot be directly linked to real users or to an identity\cite{antonopoulos2017mastering}. A transaction, thus, does not identify the payer or the payee in any way. However, by exploiting the properties of the protocol, like the fact that the transaction history is publicly available, it is possible to trace and cluster addresses that are owned by the same user, collapsing in that way a network of addresses into a network of users.
The principle that drove our approach is to minimise as much as possible the number of false positives, that is, the addresses that are linked together as if they were owned by the same user but they are not. The approach is based on two heuristics introduced by Meiklejohn \textit{et al.}\cite{meiklejohn2013fistful}, that we describe here.

\textit{Input-based Heuristic}: The first and safest one exploits the fact that, in principle, if two addresses are input of the same transaction, then they are controlled by the same user. This property is also transitive, which means that if a transaction includes in the input the addresses A and B, and a second transaction includes the addresses B and C, then it is safe to assume that A,B,C addresses belong to the same user.

\textit{One-time change addresses}: the second one detects addresses appearing in the output of a transaction and that can be attributed to the owner of the inputs.

In our work we adopt the first heuristic, but we modify the second one, as in the working paper by Tessone\cite{Tessone:2018}, with the aim of using the second heuristic only in the case in which it is safe to assume that the one-time change address belongs to the owner of the inputs.

\begin{figure}
\centering
\includegraphics[width=.5\linewidth]{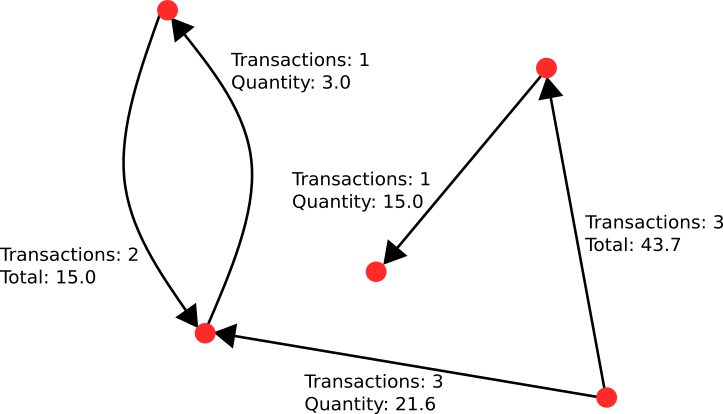}
\caption{\iac{An illustrative example of the construction of the network. Nodes, corresponding to users, are connected by directed edges, corresponding to transactions. Edges are annotated with the total amount of transactions occurred and the total quantity of bitcoins transferred during the observation window.}}
\label{fig:example}
\end{figure}

Others techniques use different approaches\cite{androulaki2013evaluating,tasca2018evolution,harrigan2016unreasonable,ron2013quantitative}, where the one-time change addresses are defined using similar heuristics. Another approach, proposed by Cazabet et al.\cite{remy2017tracking}, is based on a combination of the input-based heuristic and of the Louvain community detection algorithm to detect addresses that are likely to belong to the same user.
The reason why we did not use these methods is that we privileged the heuristics preserving the analysis from errors in detecting addresses not belonging to the same users as if they were controlled by a single entity, i.e. avoiding false positives.\\
Note that these techniques do not allow to reproduce the perfect network of users, since real users can use different wallets that not necessarily are linked together by a transaction; thus, the reader should not consider the network obtained as a perfect representation of the real network of users, but as an approximation that clusters addresses minimizing the presence of false positives.\\
Once the addresses are grouped by wallet, we build the network in the following way: two nodes (wallets) i and j are connected via a directed edge (i,j) if at least one transaction from i to j occurs during the considered time integration window. Edges are annotated with the number of transactions occurred and the total quantity of bitcoins transferred.
The structure of the network is illustrated with an example in Figure \ref{fig:example}.

\subsection*{Null models}

In order to verify the statistical significance of our results we have compared them with a properly-defined null model. Inspired by the empirical regularities of the degree distribution we have employed the Directed Random Grapg Model (DRGM). It is an Exponential Random Graph Model (ERGM) defined within the constrained Shannon entropy-maximization framework. Briefly speaking, one solves the following problem

\begin{equation}
\max_{\mathbf{P}}\left\{S[P]-\sum_i\theta_i\left[\sum_{\mathbf{A}}P(\mathbf{A})X(\mathbf{A})-\bar{X_i}\right]\right\}
\end{equation}
where the vector of constraints reads $\vec{C}(\mathbf{A})=\{\vec{k}^{out},k^{in}\}$ and $C_0=\langle C_0\rangle=1$ sums up the normalization condition of the searched probability distribution. The solution to the problem above reads

\begin{equation}
P(A)=\frac{e^{-H(\mathbf{A},\vec{\theta})}}{Z(\vec{\theta})}
\end{equation}
with $H(\mathbf{A},\vec{\theta})=\vec{\theta}\cdot\vec{C}$ summing up the proper topological constraints. In the DRGM case, our Hamiltonian reads
with $H(\mathbf{A},\theta)\equiv\theta L$ a position leading to the probability function

\begin{equation}
P(\mathbf{A})=p^L(1-p)^{N(N-1)-L}
\end{equation}
with $p\frac{e^{-\theta}}{1+e^{-\theta}}\equiv\frac{x}{1+x}$. The comparison between observed and expected properties on the ensemble has been carried out by employing the z-score index, defined as

\begin{equation}
z_X=\frac{N_X-\bar{X}}{\sigma_X};
\label{eq:zscore}
\end{equation}

in our case, $X$ represents the abundance of the triadic motifs shown in fig. \ref{fig:motifs}. The \textit{z-score} is a standardised variable measuring the difference between the observed and the expected value in units of standard deviation. If X is normally distributed under the null model, then values within $z = \pm1$, $z = \pm2$, $z = \pm3$ would (approximately) occur with a 68\%, 95\%, 99\% probability respectively. If the observed value of X corresponds to a large positive (negative) value of $z_X$ then the quantity X is over(under)-represented in the data, and not explained by the null model. 

A simpler analysis discounting the increasing volume of the network is obtained by considering the index $\hat{N}_m=\frac{N_m}{N}$, i.e. by dividing the abundance of a given motif $m$ by the total number of nodes in a particular snapshot.

\subsection*{Motifs detection}

\begin{figure}
\centering
\includegraphics[width=.95\linewidth]{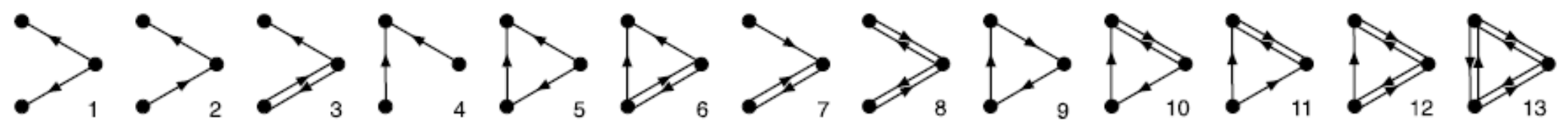}
\caption{The 13 possible triadic motifs involving three connected vertices\cite{Squartini:2013}.}
\label{fig:motifs}
\end{figure}

An exact counting of the network motifs present in the reconstructed networks of transactions was performed, on a reduced set of 17 time points in the period of the first bitcoin bubble. The 13 three-node network motifs analysed are the same ones as those described by Squartini et al.\cite{Squartini:2013} and are represented in Fig. \ref{fig:motifs}. Null models are built for each time point with the procedure described above. The expected number of network motifs $\bar{X}$ and their standard deviation $\sigma_X$ are henceforth obtained, allowing us to calculate the \textit{z-score} (equation \ref{eq:zscore}).

\section*{Acknowledgements}

All authors are grateful to Alberto Antonioni, Eugenio Valdano and IMT Lucca for the organisation of the workshop Complexity 72H during which this research was performed. C.J.T. acknowledges financial support of the University of Zurich through the University Research Priority Programme on Social Networks. V.R. was sponsored by the U.S. Army Research Laboratory and the U.K. Ministry of Defence under Agreement Number W911NF-16-3-0001. The views and conclusions contained in this document are those of the authors and should not be interpreted as representing the official policies, either expressed or implied, of the U.S. Army Research Laboratory, the U.S. Government, the U.K. Ministry of Defence or the U.K. Government. The U.S. and U.K. Governments are authorised to reproduce and distribute reprints for Government purposes notwithstanding any copyright notation herein.



\bibliographystyle{naturemag-doi}
\bibliography{biblio}

\end{document}